\documentclass[11pt]{article}
\usepackage[margin=1in]{geometry}
\usepackage[T1]{fontenc}
\usepackage[utf8]{inputenc}
\usepackage{lmodern}
\usepackage{microtype}
\usepackage{amsmath,amssymb}
\usepackage{booktabs}
\usepackage{array}
\usepackage{enumitem}
\usepackage{tikz}
\usetikzlibrary{positioning,arrows.meta,fit,backgrounds,calc}
\usepackage[hidelinks]{hyperref}

\tikzset{
  box/.style={draw, rounded corners=2pt, align=center, font=\small,
              minimum height=8mm, inner sep=4pt, fill=white},
  store/.style={draw, align=center, font=\small, minimum height=8mm,
                inner sep=4pt, fill=gray!10},
  arr/.style={-{Stealth[length=2mm]}, thick},
  lbl/.style={font=\scriptsize, fill=white, inner sep=1pt}
}

\title{Detecting Security Threats from Logon Data:\\
A Technical Framework for Baseline-Relative Anomaly Detection\\
with Expert-Derived Threat Tagging in SaaS Systems}

\author{Naga Lakshmi Subha Pavan Kumar Ghantasala \and Saibala Lakkaraju \and
Srinivas Chava \and Kristine Wetch \and Moshe Adam Van Berg \and
Elvis Dang \and Vamsi Kadiyala \and Satish Raj Katakam%
\thanks{The work described was performed at Wells Fargo.}}

\date{}

\begin{document}
\maketitle

\begin{abstract}
Organizations that run software-as-a-service (SaaS) systems expose many
applications and data repositories behind credentials they manage, and the login
traffic to those services is one of the earliest places an attack becomes
visible. Rule-based monitoring of that traffic struggles to keep pace with new
attack vectors, and slow, low-volume attacks are designed to stay below the
thresholds that rules depend on. We describe a technical framework for
authentication threat detection that combines four elements. We present it as an
architecture and a fully specified reference method, not as a detection system
whose effectiveness we have demonstrated. First, it builds a baseline of
logon behavior for each user, each category of user and each device from
authentication attributes such as time, device, user agent, application,
authentication method, request type and outcome, and it compares each user with
their peers. Second, an unsupervised model, such as an isolation forest, scores
new authentication activity against those baselines without requiring labels.
Third, subject matter experts review the resulting outliers and write tagging
rules that map them to threat categories, including brute force, credential
checking, reconnaissance, low-and-slow unauthorized access and misconfiguration,
while organizational context filters out activity that is known to be expected.
Fourth, the rule-labeled data trains a supervised classifier that runs in tandem
with the unsupervised model. The classifier has an explicit reject option, so
activity that scores as anomalous but that it cannot confidently assign to a
known category is still routed to response and expert review instead of being
forced into a category. Each outcome drives remediation through a security
information and event management (SIEM) system, control of downstream systems,
or reporting whose urgency follows the severity of the category. The framework
is disclosed in US Patent Application Publication US~2026/0006036~A1. Beyond the
publication, this paper contributes a formal statement of the method, concrete
procedures for the steps the publication leaves open (feature construction,
threshold selection, the classifier's reject option and baseline updating), a
comparison with prior work on authentication anomaly detection, user and entity
behavior analytics (UEBA) and analyst-in-the-loop learning, an evaluation
protocol on public authentication data, and a proof of concept on the Los Alamos
authentication data. In that experiment the framework's detector scored below
two simple baselines on real red-team activity (ROC-AUC 0.73 against 0.76). On
synthetic attacks injected into real traffic, the categorization stage reached a
macro-$F_1$ of 0.88 once it had enough examples, but the reject option caught most
of one held-out category and little of the others. We therefore make no claim
that the framework detects attacks better than existing approaches, including
previously unseen attacks.
\end{abstract}

\section{Introduction}
\label{sec:intro}

Cloud computing delivers computing services over a network, and SaaS providers
offer those services with convenience, flexible resources and economies of scale
\cite[\P0002]{patent}. An organization that operates a SaaS system, for example a
financial institution that runs it for employees across many physical locations,
exposes a number of services such as applications and data repositories, manages
the credentials of users and software components that access them, and monitors
login activity across the system to keep it secure \cite[\P0020]{patent}.

Authentication traffic is an attractive place to look for attacks, because most
attacks against a SaaS system must at some point present credentials. It is also
a hard place to look. As the attack surface of publicly accessible SaaS
applications grows, keeping up with the latest attack vectors through rule-based
monitoring becomes difficult \cite[\P0014]{patent}. Rules encode what defenders
already know, they must be written and maintained by hand, and they rely on
thresholds. Attackers who know this design attacks that ``fly under the radar,''
making slow and precise attempts to gain access while staying below the activity
levels required to trigger typical security systems \cite[\P0005]{patent}. The
complexity of a large SaaS system also means that traditional techniques may be
unable to identify the full range of potential issues \cite[\P0005]{patent}.

Unsupervised anomaly detection \cite{chandola} addresses part of the problem.
It needs no labeled attacks, it can surface behavior that no rule anticipated,
and detecting an anomaly lets administrators act to counter a potential threat
even when its precise nature is unknown \cite[\P0016]{patent}. On its own,
however, an unsupervised detector leaves several problems open. Its output is an
outlier score, not an explanation, so analysts must work out what each alert
means. It cannot tell an attack from an unusual but legitimate business process.
And it does not improve from the judgments analysts make about its alerts.

This paper describes a technical framework that is designed to address these
gaps by placing human expertise between unsupervised detection and supervised
classification. The framework develops a baseline of logon and related activity
and uses it to detect anomalous logons from authentication traffic
\cite[\P0014]{patent}. Subject matter experts (SMEs) map the anomalies that the
unsupervised model surfaces to identifiable threats and write tagging rules that
label similar activity when it recurs; the labeled data then trains a supervised
model that predicts known patterns \cite[\P0017]{patent}. Because the
unsupervised model keeps running alongside the supervised one, activity that no
rule or class covers can still be acted on, provided it is anomalous with respect
to the baseline features \cite[\P0018]{patent}.

\paragraph{Scope of this paper.} We present the framework as an architecture and
a reference method for authentication threat detection. We do not present it as a
detection system whose effectiveness has been demonstrated. We report one
proof-of-concept experiment on public data, whose detection results are
negative. The framework is disclosed in a US patent application
publication \cite{patent}, whose inventors are the authors of this paper. That
document describes the components and their interactions but leaves several
procedures open, and it does not position the method against the research
literature. Where we state what the publication describes, we cite its
paragraphs; where we specify something it leaves open, we say so.

\paragraph{Contributions.} Beyond the publication and existing approaches, this
paper makes the following contributions.
\begin{enumerate}[leftmargin=*,itemsep=2pt]
\item \emph{A formal statement of the framework.} We define the entities,
baselines, scores, tagging rules and decision logic precisely
(Sections~\ref{sec:problem}--\ref{sec:tandem}), so that independent
implementations can be built and compared.
\item \emph{Concrete procedures for the open steps.} We specify the event
features (Section~\ref{sec:features}), threshold selection under an analyst
alert budget (Section~\ref{sec:threshold}), a reject option for the supervised
classifier that combines a confidence test with a distance test
(Section~\ref{sec:reject}), and a baseline-updating procedure with cold-start
handling and quarantine of flagged events (Section~\ref{sec:update}). None of
these is fixed by the publication.
\item \emph{A positioning against prior work.} We compare the framework with
authentication anomaly detection, UEBA and analyst-in-the-loop systems, in
particular AI$^2$ \cite{ai2} and Snorkel \cite{snorkel}, and state which design
elements differ (Section~\ref{sec:related}, Table~\ref{tab:compare}).
\item \emph{An evaluation protocol.} We define tasks, metrics, baselines and
ablations on the public Los Alamos authentication data \cite{lanl}, including a
leave-one-category-out test of the reject option (Section~\ref{sec:eval}).
\item \emph{A proof of concept.} We implement a reduced form of the framework
on the Los Alamos data and report its results against simple baselines, including
where it falls short (Section~\ref{sec:poc}).
\end{enumerate}
The individual components, such as isolation forests, peer-group comparison,
rule-based labeling and classification with a reject option, are known. What we
contribute is their combination for authentication data, the specification of
how they interact, and a protocol for testing whether the combination helps.

The rest of the paper is organized as follows. Section~\ref{sec:related} reviews
related work. Section~\ref{sec:problem} defines the setting and threat model.
Section~\ref{sec:arch} describes the system architecture.
Sections~\ref{sec:data}--\ref{sec:response} describe each stage of the method,
and Section~\ref{sec:e2e} walks through its end-to-end operation.
Section~\ref{sec:scenarios} illustrates the framework on representative
scenarios, Section~\ref{sec:eval} sets out the evaluation protocol,
Section~\ref{sec:poc} reports the proof of concept,
Section~\ref{sec:discussion} discusses design choices and limitations, and
Section~\ref{sec:conclusion} concludes. Appendix~\ref{app:claims} maps the
claims of the publication to the sections of this paper.

\section{Related Work}
\label{sec:related}

\paragraph{Unsupervised anomaly detection.} Anomaly detection has a long history
in settings where labels are scarce \cite{chandola}. Density- and
neighborhood-based methods such as the local outlier factor \cite{lof}, and
boundary methods such as the one-class support vector machine \cite{ocsvm}, model
normal data and flag points that fall outside it. Isolation forests
\cite{iforest,iforest2} take the opposite approach and isolate anomalies
directly with random partitions, which makes them fast, insensitive to feature
scaling and well suited to the mixed, high-volume features found in
authentication logs. Our framework uses an isolation forest in some examples and
treats clustering, association-rule and dimensionality-reduction models as
alternatives \cite[\P0028, \P0076]{patent}.

\paragraph{Authentication anomaly detection.} A large body of work detects
malicious logins from authentication logs. Freeman et al.\ \cite{freeman} score
the authenticity of a login from the history of attributes such as IP address
and user agent, and show that such scores can strengthen password
authentication; risk-based authentication deployed in practice follows the same
idea \cite{wiefling}. A second line of work models the structure of logins
inside an enterprise network. Kent et al.\ \cite{kentgraphs} represent each
user's authentications as a graph and show that graph properties characterize
user behavior. Siadati and Memon \cite{siadati} learn the structural patterns of
legitimate logins between machines and flag logins that fit none of them.
Bohara et al.\ \cite{bohara} combine several unsupervised detectors over host
logs to identify lateral movement. Powell \cite{powell} detects malicious logins
as anomalies in login graphs, Log2vec \cite{log2vec} embeds heterogeneous log
graphs to detect threats, Euler \cite{euler} detects lateral movement by
temporal link prediction, and Hopper \cite{hopper} reconstructs login paths to
detect lateral movement with few false positives. Several of these evaluate on
the Los Alamos data \cite{lanl}. These methods produce strong detectors for
specific behaviors, mostly lateral movement, but they output an anomaly score or
a binary decision. They do not attach a threat category to an alert, and they do
not describe how analyst judgments flow back into the system.

\paragraph{User and entity behavior analytics.} UEBA systems baseline the
behavior of users and entities and flag deviations. Tuor et al.\ \cite{tuor}
train recurrent and feed-forward networks on per-user activity streams for
unsupervised insider threat detection; more recent work applies deep
autoencoders to enterprise logs \cite{fuentes} and to identity-provider logs
\cite{cain}, and uses large language models to classify endpoint security logs
\cite{henshaw}. Surveys of behavior-based anomaly detection
\cite{akampurira,lanuwabang} and of deep-learning intrusion detection \cite{xu}
identify the scarcity of labeled attacks and the difficulty of interpreting model
outputs as recurring obstacles. Our baselines per user, role, device and context
and our peer comparison follow the UEBA tradition; the framework differs in what
happens after an event is scored.

\paragraph{Analysts in the loop.} Beehive \cite{beehive} mines enterprise logs
with unsupervised clustering and hands the resulting incidents to analysts for
investigation. AI$^2$ \cite{ai2} ranks outliers from an ensemble of unsupervised
detectors, shows the top-ranked events to analysts, and uses their labels to
train a supervised model that is combined with the detectors in later rounds.
Active anomaly discovery \cite{das} uses expert feedback on each presented
anomaly to re-weight the detector so that later queries are more likely to be
true anomalies. Weak supervision frameworks such as Snorkel \cite{snorkel} let
experts write labeling functions instead of labeling examples one by one, and
combine the noisy, overlapping outputs of those functions with a generative
label model.

\paragraph{How the framework differs from AI$^2$ and Snorkel.} The framework is
closest to AI$^2$, and its tagging rules play the role of Snorkel's labeling
functions. It differs from both in four ways.
\begin{enumerate}[leftmargin=*,itemsep=1pt]
\item \emph{Labels are categories, not verdicts.} AI$^2$ collects
malicious-or-benign labels. Here experts write rules that assign a named threat
category, so the supervised model is multi-class and its output selects a
category-specific response (Section~\ref{sec:response}).
\item \emph{Rules are written from, and applied to, detected anomalies.} Snorkel
applies labeling functions to a whole unlabeled corpus. Here rules are written by
inspecting the detector's outliers and are applied to events the detector flags,
so they only need to separate threat categories within anomalous activity, not
separate attacks from all normal traffic.
\item \emph{Abstention is an operational outcome.} In Snorkel, examples on which
every labeling function abstains are left out of training. In AI$^2$ the
supervised model's output is combined with the detectors into a single ranking.
Here the case in which no rule fires, or the classifier rejects, is kept as a
distinct outcome that triggers a response and is queued for expert review as a
candidate new category (Section~\ref{sec:tandem}).
\item \emph{Organizational context is an explicit filter.} Known business
activity is removed by an explicit context step when baselines are formed and
when rules are written (Section~\ref{sec:context}), rather than learned
implicitly from benign labels.
\end{enumerate}
Table~\ref{tab:compare} summarizes these differences. The comparison is coarse
and is based on the designs described in the cited papers, not on experiments.

\begin{table}[t]
\centering
\caption{Design elements of representative approaches. A check means that the
design includes the element as described in the cited work. The
comparison is qualitative.}
\label{tab:compare}
\small
\begin{tabular}{@{}>{\raggedright}p{0.33\linewidth}ccccc@{}}
\toprule
Approach & \shortstack{Unsup.\\scoring} & \shortstack{Expert\\feedback} &
\shortstack{Named\\categories} & \shortstack{Supervised\\model} &
\shortstack{Unclassified\\path} \\
\midrule
Graph-based login detection \cite{powell,log2vec,euler} & $\checkmark$ & & & & \\
Autoencoder UEBA \cite{fuentes,cain} & $\checkmark$ & & & & \\
Active anomaly discovery \cite{das} & $\checkmark$ & $\checkmark$ & & & \\
AI$^2$ \cite{ai2} & $\checkmark$ & $\checkmark$ & & $\checkmark$ & \\
Snorkel \cite{snorkel} & & $\checkmark$ & $\checkmark$ & $\checkmark$ & \\
This framework & $\checkmark$ & $\checkmark$ & $\checkmark$ & $\checkmark$ & $\checkmark$ \\
\bottomrule
\end{tabular}
\end{table}

\paragraph{Classification with a reject option.} Deciding when a classifier
should decline to predict is a classical problem. Chow \cite{chow} derived the
optimal trade-off between error and rejection from posterior probabilities;
selective classification \cite{geifman} controls the risk of the accepted
predictions; and open-set recognition \cite{scheirer} addresses test inputs from
classes absent in training, for which a confident softmax output is unreliable
\cite{hendrycks}. Our reject option (Section~\ref{sec:reject}) draws on both: a
confidence test in the spirit of Chow's rule and a distance test against the
training data of the predicted category, in the spirit of open-set recognition.

\paragraph{Threat categories and the limits of learning.} Threat taxonomies such
as MITRE ATT\&CK \cite{attack} organize adversary behavior into tactics and
techniques that defenders can map detections to. Sommer and Paxson
\cite{sommer} argue that machine learning for intrusion detection is harder than
in many other domains, because attacks are rare, the cost of errors is high, and
alerts must be explained to operators. These observations motivate both the use
of named threat categories and the human review stage of our framework, and they
are why we make no claim of effectiveness without evaluation.

\section{Problem Setting and Threat Model}
\label{sec:problem}

\subsection{Setting}

We consider a network that logically connects a set of services $\mathcal{S}$,
such as SaaS applications, database access services and other cloud services,
and that provides access to them through a single point of contact
\cite[\P0022, \P0023]{patent}. The services are supported by infrastructure
nodes, which include hardware and software network components such as the
network management software that manages access to the services
\cite[\P0024]{patent}. External devices reach the network through a gateway made
up of routers, switches and management software \cite[\P0025]{patent}. Users
operate devices such as laptops, desktops, smartphones, tablets and augmented or
virtual reality headsets, and some users are automated services that interact
with the services autonomously \cite[\P0036]{patent}. Each device authenticates
to a service by sending an authentication request that carries credentials, and
the service decides whether to grant access \cite[\P0037, \P0038]{patent}.

Table~\ref{tab:notation} summarizes our notation. An authentication event
$e = (u, d, s, t, \mathbf{a})$ records that user $u$ on device $d$ requested
access to service $s$ at time $t$, with attributes $\mathbf{a}$ described in
Section~\ref{sec:data}. Each user belongs to one or more categories $r \in
\mathcal{R}$, such as a role in the organization. Historical activity is the
record of events before a point in time, and recent activity is the set of
events within a predetermined recent period, such as the last hour, day or week
\cite[\P0027]{patent}.

\begin{table}[t]
\centering
\caption{Notation.}
\label{tab:notation}
\small
\begin{tabular}{@{}ll@{}}
\toprule
Symbol & Meaning \\
\midrule
$\mathcal{U}, \mathcal{D}, \mathcal{S}$ & Users (including automated services), devices, services \\
$\mathcal{R}$ & Categories of users, such as roles \\
$e=(u,d,s,t,\mathbf{a})$ & Authentication event with attributes $\mathbf{a}$ \\
$W_t$ & Baseline window ending at time $t$, for example 14 days \\
$\mathbf{b}_x(W_t)$ & Baseline metrics of entity $x$ over $W_t$ \\
$G(u)$ & Peer group of user $u$ \\
$\mathbf{x}_e$ & Feature vector of event $e$ \\
$s_U(e)$ & Anomaly score from the unsupervised model \\
$\tau$ & Anomaly threshold \\
$\mathcal{C}$ & Set of threat categories \\
$\hat{c}(e)$ & Category assigned by tagging rules or the supervised model \\
\bottomrule
\end{tabular}
\end{table}

\subsection{Objectives}

Given the stream of authentication events and related network activity, the
system must (i) decide whether each event, or a set of related events, is
anomalous relative to the baseline of network activity; (ii) classify anomalous
activity into a threat category from a set $\mathcal{C}$, or determine that it
fits none; and (iii) take an action, based on the category, that mitigates the
threat \cite[\P0006]{patent}. The set $\mathcal{C}$ is not fixed. It contains
predetermined categories, derived from known types of malicious activity and
public databases of attack vectors and techniques, and categories that the system
identifies from novel attack vectors surfaced by the unsupervised model
\cite[\P0059]{patent}.

\subsection{Threat model}

The framework targets threats that reveal themselves through authentication and
the activity that follows it. The risks the publication names are brute force
attacks on services and nodes, scanning for compromised credentials or attempts
to use them (credential checking), reconnaissance of services and nodes,
unauthorized access through low-and-slow attacks designed to avoid attack
detection systems, and misconfigurations in the network
\cite[\P0097]{patent}. A device that has gained access to one service may then
probe others, so the framework also observes operations on the nodes that
support the services \cite[\P0043]{patent}. Accounts with privileges that allow
access to more sensitive details deserve particular attention, because the
types of information an account accesses can characterize unusual behavior or
reveal sensitive resources that had not been considered targets
\cite[\P0043]{patent}. Misconfiguration is included deliberately: it is not an
attack, but it produces anomalous authentication patterns and creates openings
that attackers can use. We assume that the collector can observe authentication
requests and node activity, and that most activity in a baseline window is
benign, an assumption we revisit in Section~\ref{sec:discussion}.

\section{System Architecture}
\label{sec:arch}

\begin{figure}[t]
\centering
\begin{tikzpicture}
  \node[box, minimum width=24mm] (dev) at (-0.5,0) {User devices\\(people and\\automated services)};
  \node[box, minimum width=18mm] (gw) at (3.6,0) {Gateway};
  \node[box, minimum width=28mm] (svc) at (7.4,0.8) {Services (SaaS apps,\\data repositories)};
  \node[box, minimum width=28mm] (nodes) at (7.4,-0.9) {Infrastructure nodes};
  \begin{scope}[on background layer]
    \node[draw, dashed, rounded corners, fit=(gw)(svc)(nodes), inner sep=3mm,
          label={[font=\small]above:Network}] (net) {};
  \end{scope}
  \node[box, minimum width=36mm] (an) at (7.4,-3.6) {Analysis system\\[1pt]
        \scriptsize collector $\cdot$ unsupervised model\\
        \scriptsize supervised model};
  \node[box, minimum width=24mm] (sme) at (1.8,-3.6) {SME system};
  \node[box, minimum width=24mm] (mdev) at (1.8,-5.6) {Model development\\systems};
  \node[box, minimum width=26mm] (rem) at (12.4,-3.6) {Remediation\\system (e.g., SIEM)};
  \node[box, minimum width=26mm] (ctl) at (12.4,-5.6) {Controlled\\systems};

  \draw[arr] (dev) -- node[lbl, above=2pt]{requests} (gw);
  \draw[arr] (gw.east) -- (svc.west);
  \draw[arr] (svc.east) -- ++(6mm,0) |- node[lbl, pos=0.12, right=1pt]{login data} ([yshift=2.5mm]an.east);
  \draw[arr] (nodes.south) -- node[lbl]{network data} (an.north);
  \draw[arr, {Stealth[length=2mm]}-{Stealth[length=2mm]}] (sme) -- node[lbl, above=2pt]{rules, feedback} (an);
  \draw[arr] (mdev.east) -| node[lbl, pos=0.25, above=2pt]{models, rules} (an.south);
  \draw[arr] ([yshift=-2.5mm]an.east) -- node[lbl, below=3pt]{control signal} ([yshift=-2.5mm]rem.west);
  \draw[arr] (rem) -- node[lbl, right=2pt]{control signal} (ctl);
\end{tikzpicture}
\caption{System architecture, after FIGS.~1A and~1B of the publication
\cite{patent}. A service may also consult the anomaly score when it decides
whether to approve a request (Section~\ref{sec:unsup}).}
\label{fig:arch}
\end{figure}
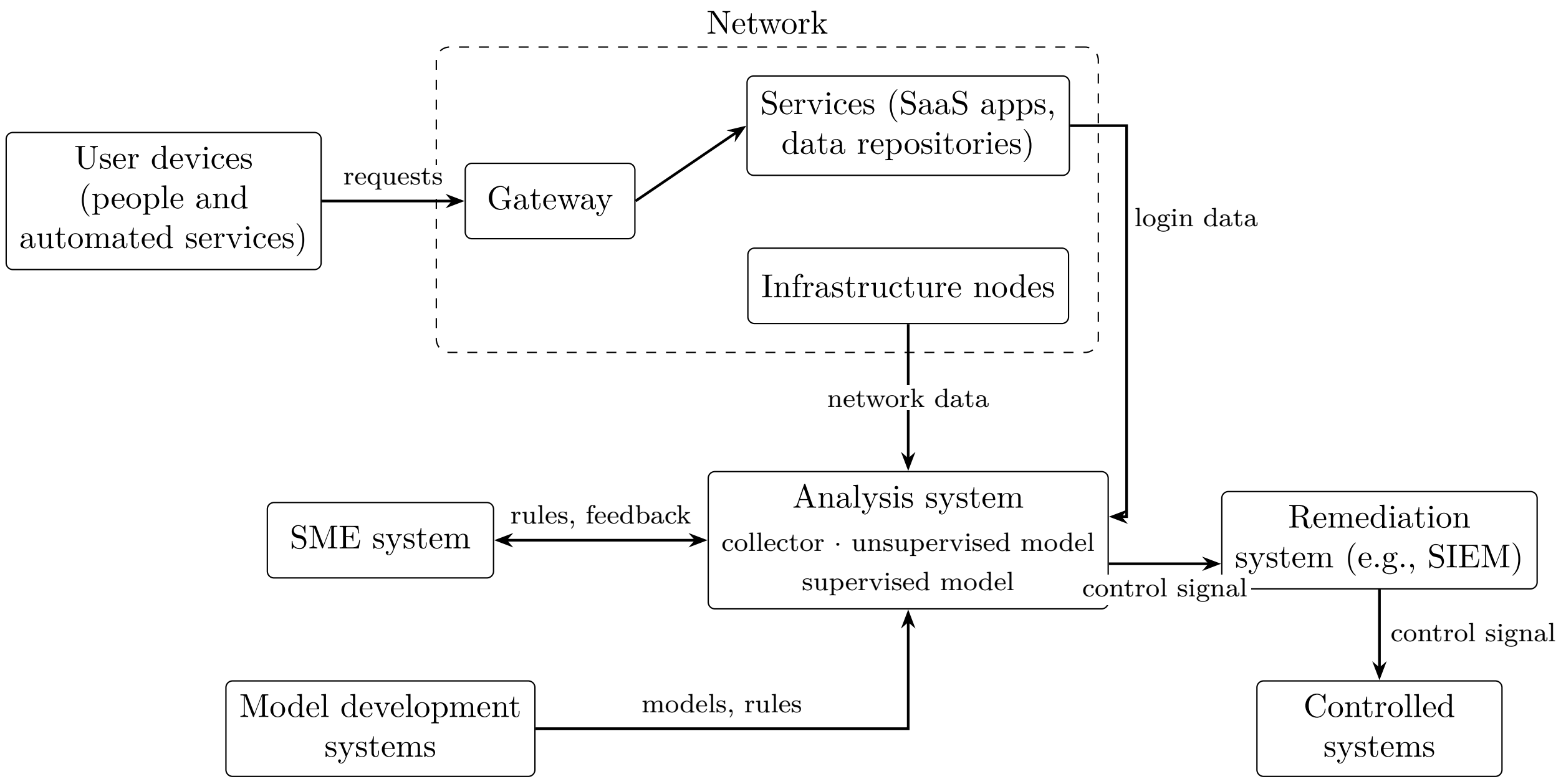

Figure~\ref{fig:arch} shows the components of the framework and the flows
between them. Users' devices send authentication requests through the gateway,
which routes each request to the service it is destined for; the service
evaluates it and grants or denies access \cite[\P0038--\P0040]{patent}. An
analysis system collects information about this activity and analyzes it with
one or more models, including an unsupervised model and a supervised model
\cite[\P0026, \P0028, \P0029]{patent}. Three further systems surround it. An SME
system lets subject matter experts give feedback on tagging rules, supervised
models and other domain-specific topics, including modifications of tagging
rules and annotations of activity for training \cite[\P0034]{patent}. Model
development systems, operated by developers and data scientists, manage the
development, training and calibration of the models and administer the creation
and application of tagging rules \cite[\P0033]{patent}. A remediation system,
which may be a SIEM system, acts on the insights the analysis system produces,
helps analysts spot threats that might otherwise go undetected, and helps the
organization meet reporting requirements \cite[\P0031]{patent}. It may in turn
send control signals to controlled systems downstream \cite[\P0032]{patent}.

\subsection{The analysis system}

Figure~\ref{fig:modules} and Table~\ref{tab:modules} describe the modules of
the analysis system, following FIG.~2 of the publication. A collector module
obtains historical and recent network activity and stores it in a collector data
store such as a database or data lake \cite[\P0072, \P0073]{patent}. An analysis
module orchestrates the work: it asks the collector for activity more recent than
what is stored, has the models identify and classify anomalous activity, and
engages the remediation module to act on the result \cite[\P0074]{patent}. The
models comprise unsupervised models and supervised models
\cite[\P0075]{patent}. A model development module captures the anomaly scores
of the unsupervised models and processes them into rules, labels and
annotations, using information from SMEs held in an SME data store, and develops
the supervised models from the labeled data \cite[\P0079--\P0081]{patent}. A
context data store holds information about the organization that manages the
network, the usage characteristics of the network and its organizational
context, such as business practices, employee types and normal usage patterns
\cite[\P0082]{patent}. A remediation module acts on classified activity
\cite[\P0085]{patent}, and a reporting module produces reports and alerts
\cite[\P0086]{patent}.

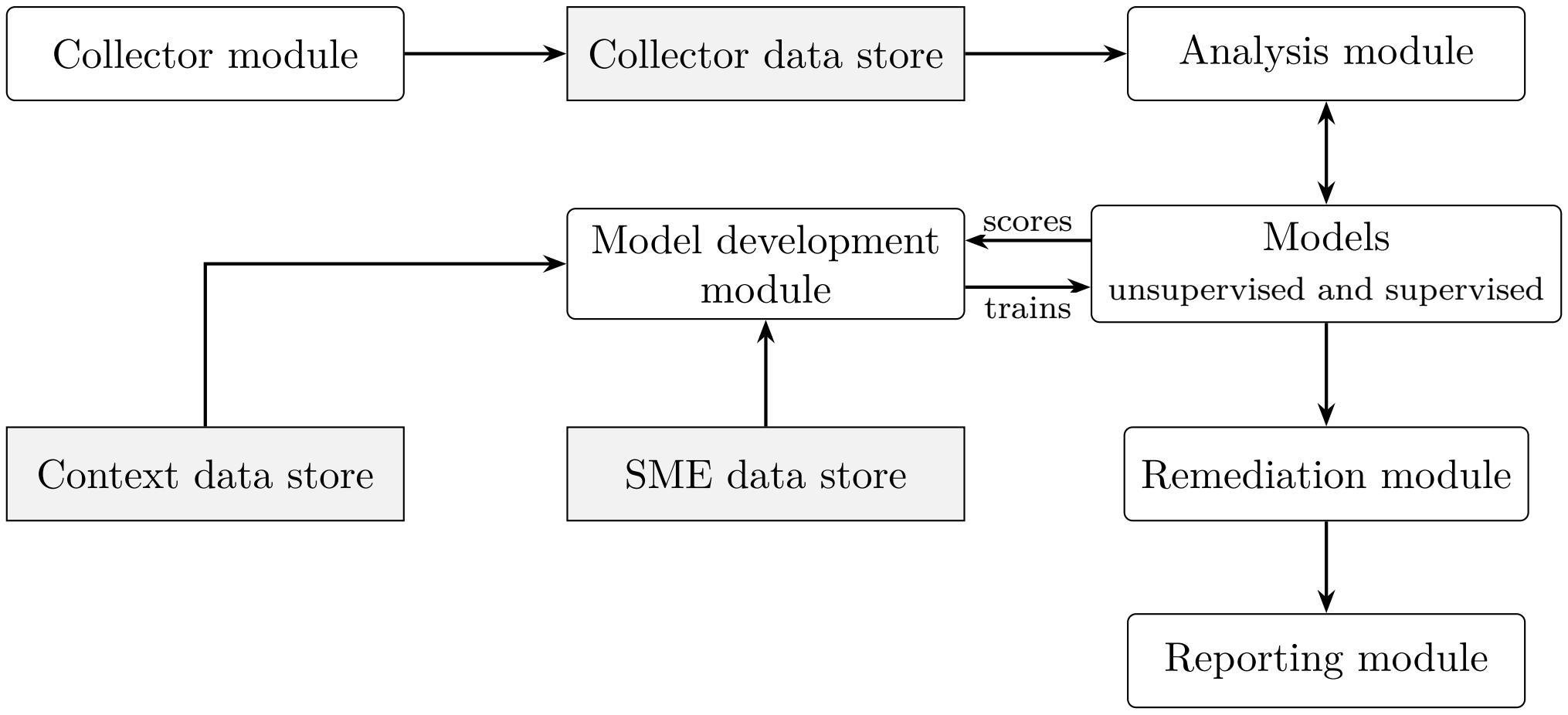
\begin{figure}[t]
\centering
\begin{tikzpicture}
  \node[box, minimum width=34mm] (col) at (0,0) {Collector module};
  \node[store, minimum width=34mm] (cds) at (4.8,0) {Collector data store};
  \node[box, minimum width=34mm] (ana) at (9.6,0) {Analysis module};
  \node[box, minimum width=34mm] (mdm) at (4.8,-1.8) {Model development\\module};
  \node[box, minimum width=34mm] (mod) at (9.6,-1.8) {Models\\\scriptsize unsupervised and supervised};
  \node[store, minimum width=34mm] (ctx) at (0,-3.6) {Context data store};
  \node[store, minimum width=34mm] (smes) at (4.8,-3.6) {SME data store};
  \node[box, minimum width=34mm] (rem) at (9.6,-3.6) {Remediation module};
  \node[box, minimum width=34mm] (rep) at (9.6,-5.2) {Reporting module};
  \draw[arr] (col) -- (cds);
  \draw[arr] (cds) -- (ana);
  \draw[arr, {Stealth[length=2mm]}-{Stealth[length=2mm]}] (ana) -- (mod);
  \draw[arr] ([yshift=2mm]mod.west) -- node[lbl, above=1pt]{scores} ([yshift=2mm]mdm.east);
  \draw[arr] ([yshift=-2mm]mdm.east) -- node[lbl, below=1pt]{trains} ([yshift=-2mm]mod.west);
  \draw[arr] (smes) -- (mdm);
  \draw[arr] (ctx.north) |- (mdm.west);
  \draw[arr] (mod) -- (rem);
  \draw[arr] (rem) -- (rep);
\end{tikzpicture}
\caption{Modules of the analysis system, after FIG.~2 of the publication
\cite{patent}. Shaded boxes are data stores.}
\label{fig:modules}
\end{figure}

\begin{table}[t]
\centering
\caption{Responsibilities of the analysis system's modules.}
\label{tab:modules}
\small
\begin{tabular}{@{}>{\raggedright}p{0.26\linewidth}>{\raggedright\arraybackslash}p{0.66\linewidth}@{}}
\toprule
Module & Responsibility \\
\midrule
Collector module & Obtains login data from services and network data from nodes, historical and recent \\
Collector data store & Keeps the record of historical and recent network activity \\
Analysis module & Orchestrates collection, scoring, classification and remediation \\
Unsupervised models & Score activity against baselines; outliers indicate anomalous activity \\
Supervised models & Classify activity into threat categories, trained on rule-labeled data \\
Model development module & Turns anomaly scores and SME input into rules, labels, annotations and categories; trains supervised models \\
SME data store & Keeps SME feedback on scores, classifications, rules and annotations \\
Context data store & Keeps organizational context: business practices, employee types, normal usage \\
Remediation module & Takes actions such as applying a security patch to a targeted service \\
Reporting module & Produces periodic reports and, for severe categories, immediate alerts \\
\bottomrule
\end{tabular}
\end{table}

\subsection{Deployment}

The analysis system can run on server computers, workstations, mainframes,
appliances or cloud computing systems, or on virtualized compute instances such
as virtual machines and containers in a data center or server cluster, and it can
be offered as a web service \cite[\P0030]{patent}. It is described as separate
from the services and nodes it monitors, but some or all of it may be
incorporated into those systems, including the nodes \cite[\P0030]{patent}; the
remediation system may likewise be part of the analysis system
\cite[\P0031]{patent}. The modules may be implemented in software, hardware or
firmware, combined into a single module or split across several devices
\cite[\P0087, \P0088]{patent}. This flexibility matters in practice: the same
logic can run as a batch job over a data lake, as a streaming service next to an
identity provider, or inside the network management layer.

\section{Data Collection and Logon Attributes}
\label{sec:data}

\subsection{Collection}

The collector gathers two kinds of data. Login data describes authentication
requests and how the services responded to them. Network data describes
operations on the infrastructure nodes, including activity by both authenticated
and unauthenticated devices \cite[\P0041, \P0044]{patent}. The collector can
obtain data by polling services and nodes, by scraping data from them, by having
services actively report each login they process, or by passively observing
traffic on the network \cite[\P0027, \P0041, \P0044]{patent}. In a pull mode,
the collector identifies the services and nodes as sources and sends each a
request that may specify a period of time and the type of data sought
\cite[\P0027]{patent}. Collecting both kinds of data is important: a device that
signs into one service and then tries to sign into several others in a way that
is outside the norm may be probing for vulnerabilities, and that pattern spans
the boundary between login data and node activity \cite[\P0043]{patent}.

\subsection{Logon attributes}

Table~\ref{tab:attributes} lists the attributes the framework analyzes. The
core logon attributes are the time of the request, the properties of the device
that made it, the user agent, the application type the user attempted to log in
to, the authentication method (for example multi-factor authentication,
password, keycode or physical authentication), the request type and the
authentication status \cite[\P0042, \P0093]{patent}. The collector also gathers
the IP address and location of the source device and the login success rate
\cite[\P0015]{patent}; information about web and mobile browsers, including the
browser type, version and plugins \cite[\P0091]{patent}; and information about
the devices themselves, including MAC and IP addresses, operating system type
and revision, hardware configuration and associated users, for example by
polling nodes for the devices connected to them \cite[\P0092]{patent}. The
operations a user typically performs after successful authentication complete
the picture \cite[\P0045]{patent}.

\begin{table}[t]
\centering
\caption{Logon and related attributes collected by the framework.}
\label{tab:attributes}
\small
\begin{tabular}{@{}>{\raggedright}p{0.25\linewidth}>{\raggedright}p{0.47\linewidth}>{\raggedright\arraybackslash}p{0.18\linewidth}@{}}
\toprule
Attribute & Examples & Source \\
\midrule
Time of request & Timestamp, time of day & \cite[\P0042, \P0093]{patent} \\
Device properties & Workstation or other device, operating system, hardware & \cite[\P0092, \P0093]{patent} \\
User agent and browser & Web or mobile browser, type, version, plugins & \cite[\P0091, \P0093]{patent} \\
Application type & Service or application the user authenticates to & \cite[\P0093]{patent} \\
Authentication method & MFA, password, keycode, physical authentication & \cite[\P0093]{patent} \\
Request type & Kind of login request & \cite[\P0093]{patent} \\
Authentication status & Pass or fail; success rate over time & \cite[\P0015, \P0093]{patent} \\
Network origin & Source IP address, location, MAC address & \cite[\P0015, \P0092]{patent} \\
Post-login operations & Operations typically performed after authentication & \cite[\P0045]{patent} \\
Node activity & Operations on nodes by authenticated and unauthenticated devices & \cite[\P0043, \P0044]{patent} \\
\bottomrule
\end{tabular}
\end{table}

\section{Baselines and Peer Comparison}
\label{sec:baseline}

\subsection{Entity baselines}

From historical login data and network data, the analysis system determines a
baseline of activity for specific users and specific devices, and also for
categories of users and categories of devices \cite[\P0045]{patent}. A user's
baseline covers a recent window, such as the most recent 14 days, and indicates
what is normal for that user: how the user tends to interact with the network
and how the user accesses its services \cite[\P0045]{patent}. The window can be
a week, 14 days, a month or another timeframe \cite[\P0094]{patent}. Category
baselines capture the fact that users in different roles have different usage
patterns, so marketing, finance and research personnel each receive their own
baseline \cite[\P0046]{patent}. Baselines can also be formed for the users of a
specific service, for users operating a specific device, for users accessing
services at a specific time of day, and for other contexts
\cite[\P0046]{patent}, and for types of devices, such as the devices associated
with marketing or sales operations \cite[\P0110]{patent}.

Let $W_t = [t - \Delta, t)$ be the baseline window, with $\Delta = 14$ days by
default. For an entity $x$, which may be a user, a device, a category or a
context, the framework extracts the baseline metrics of FIG.~3 of the
publication \cite[\P0094]{patent}:
\begin{equation}
\mathbf{b}_x(W_t) = \bigl(A_x,\; I_x,\; D_x,\; \rho_x,\; Q_x\bigr),
\end{equation}
where $A_x$ is the number of applications or services the entity has logged
into, $I_x$ the number of IP addresses associated with its login attempts, $D_x$
the number of devices observed using its credentials, $\rho_x$ the fraction of
attempted logins that succeeded, and $Q_x$ the number of request types observed,
all computed over $W_t$. The richer context described in the publication,
including the services accessed, the types of device used and the operations
typically performed after authentication \cite[\P0045]{patent}, can be added as
further components.

\subsection{Comparison with other users}

A baseline says what is normal for one user; it does not say whether that
normal is itself unusual. The framework therefore compares each user's activity
with the activity of other users \cite[\P0095]{patent} on six dimensions: the
number of applications observed, the number of logins and their status, the
number of IP addresses, the number of request types, the number of browsers and
the number of user agents. The publication leaves the form of the comparison
open. One natural instantiation, robust to the heavy tails of activity counts,
compares a metric $m$ for user $u$ with the same metric over a peer group
$G(u)$, such as the users with the same role:
\begin{equation}
z_m(u) = \frac{m(u) - \operatorname{med}_{v \in G(u)} m(v)}
              {1.4826\,\operatorname{MAD}_{v \in G(u)} m(v) + \epsilon},
\label{eq:peer}
\end{equation}
where MAD is the median absolute deviation and $\epsilon$ prevents division by
zero for groups with no spread. Table~\ref{tab:features} summarizes the
baseline and peer-comparison features.

\begin{table}[t]
\centering
\caption{Baseline features (stage 3 of the pipeline) and peer-comparison
features (stage 4), from FIG.~3 of the publication \cite[\P0094, \P0095]{patent}.}
\label{tab:features}
\small
\begin{tabular}{@{}>{\raggedright}p{0.44\linewidth}>{\raggedright\arraybackslash}p{0.46\linewidth}@{}}
\toprule
Baseline over the window & Compared with other users \\
\midrule
Number of applications observed & Number of applications observed \\
Number of IP addresses observed & Number of IP addresses \\
Number of devices observed & Number of logins and their status \\
Success and failure rate & Number of request types \\
Number of request types observed & Number of browsers \\
 & Number of user agents \\
\bottomrule
\end{tabular}
\end{table}

\subsection{Organizational context}
\label{sec:context}

Some activity is unusual and yet entirely expected: a quarter-end batch job, a
planned migration, a team that legitimately works across many applications. The
framework applies organizational or business context when it observes network
activity, so that known activities of the organization are less likely to be
erroneously identified as anomalous \cite[\P0019]{patent}. In effect, context
filters out noise and lets the framework focus less on known activity and more
intensely on unknown activity \cite[\P0019]{patent}. Context can be applied when
the baseline is determined, by identifying contextual information about the
organization that controls the network and taking it into account
\cite[claim~10]{patent}, and again when tagging rules, annotations and
supervised models are developed \cite[\P0081, \P0103]{patent}.

\subsection{Feature construction}
\label{sec:features}

The publication names the attributes and the baseline and peer-comparison
metrics, but not how they are turned into the input of a model. We specify the
following construction. It is one reasonable choice, not the only one.

\paragraph{Unit of analysis.} Some threats are visible in a single event, such
as a login from a device never seen for the user. Others are visible only in
aggregate, such as credential checking, in which one source tries many accounts
a few times each. We therefore score two kinds of unit: individual
authentication events $e$, and \emph{source windows} $(o, [t-\delta, t))$ that
collect all events from one source $o$, such as an IP address or a device, over
a short window $\delta$, for example one hour. Each unit has its own feature
vector and is scored by its own model.

\paragraph{Event features.} For an event $e=(u,d,s,t,\mathbf{a})$ the feature
vector $\mathbf{x}_e$ has four groups, summarized in Table~\ref{tab:eventfeat}.
\begin{enumerate}[leftmargin=*,itemsep=1pt]
\item \emph{Novelty indicators}: whether the IP address, device, user agent,
application and authentication method of $e$ appeared in the user's baseline
window $W_t$, and whether the IP address or location appeared for any member of
the user's peer group.
\item \emph{Rarity}: for each categorical attribute value $v$, the negative log
of its smoothed frequency in the user's baseline,
$-\log\bigl((n_u(v)+1)/(N_u+K)\bigr)$, where $n_u(v)$ is the count of $v$,
$N_u$ the user's event count and $K$ the number of distinct values. Time of day
is treated the same way, using the user's histogram over 24 hourly bins.
\item \emph{Recent activity against baseline}: counts over a recent window
$\delta$ ending at $t$ (attempts, failures, distinct applications, distinct IP
addresses, distinct devices), each divided by the user's average count per
window of length $\delta$ in $W_t$, plus one.
\item \emph{Peer comparison}: the six peer scores $z_m(u)$ of Equation~\eqref{eq:peer}.
\end{enumerate}

\paragraph{Source-window features.} For a source window the features are the
number of attempts, the number of distinct target users, the number of distinct
services, the failure rate, the fraction of target users for whom the source is
new, and the inter-arrival statistics of the attempts (mean and coefficient of
variation). Very regular inter-arrival times are characteristic of automated
attempts, and low counts spread evenly over long windows are characteristic of
low-and-slow activity, which is why the same features are also computed over a
long window such as one day.

\paragraph{Encoding.} All features are numeric. Categorical attributes enter
only through novelty and rarity, so no one-hot encoding of high-cardinality
fields such as IP addresses is needed, and the features do not need scaling for
an isolation forest. Missing attributes are encoded as a separate value, because
a missing user agent can itself be informative.

\begin{table}[t]
\centering
\caption{Event features in the reference instantiation. The attributes come from
the publication; the construction is ours.}
\label{tab:eventfeat}
\small
\begin{tabular}{@{}>{\raggedright}p{0.24\linewidth}>{\raggedright}p{0.48\linewidth}>{\raggedright\arraybackslash}p{0.18\linewidth}@{}}
\toprule
Group & Features & Captures \\
\midrule
Novelty & New IP, device, user agent, application or method for the user; new IP or location for the peer group & First-time access \\
Rarity & Smoothed negative log frequency of each attribute value and of the hour of day & Unusual but not new values \\
Recent vs.\ baseline & Attempts, failures, distinct applications, IP addresses and devices in the last $\delta$, relative to the baseline rate & Bursts \\
Peer comparison & Six robust peer scores $z_m(u)$ & Unusual normal \\
\bottomrule
\end{tabular}
\end{table}

\subsection{Baseline updating}
\label{sec:update}

The publication specifies a recent window, such as 14 days, but not how the
baseline is maintained as time passes. We specify four rules.

\paragraph{Rolling recomputation.} Baselines and peer statistics are recomputed
on a fixed schedule, for example daily, over the window $W_t$ ending at the time
of recomputation. Between recomputations, events are scored against the most
recent baseline.

\paragraph{Quarantine of flagged events.} An attacker who is active during the
window would otherwise be absorbed into the baseline. Events that were flagged
as anomalous are therefore excluded from baseline computation until they are
resolved. Events that an expert confirms as benign, or that organizational
context marks as expected, are returned to the baseline; events confirmed as
malicious are excluded permanently and retained as labeled examples.

\paragraph{Cold start.} Users and devices with little history have unreliable
baselines. We shrink an entity's baseline toward the baseline of its category,
\begin{equation}
\tilde{\mathbf{b}}_u = \frac{n_u}{n_u+\kappa}\,\mathbf{b}_u +
\frac{\kappa}{n_u+\kappa}\,\mathbf{b}_{r(u)},
\label{eq:shrink}
\end{equation}
where $n_u$ is the number of events of user $u$ in $W_t$, $r(u)$ is the user's
category, such as a role, and $\kappa$ sets how many events are needed before
the user's own baseline dominates. Novelty indicators are suppressed for an
entity until it has a minimum history, so that a new employee's first logins do
not all appear new.

\paragraph{Model refresh and drift.} The unsupervised model is retrained on a
slower schedule than the baselines, for example weekly, on the quarantined
window. A shift in the distribution of anomaly scores between consecutive
periods, measured for example by the population stability index, triggers
recalibration of the threshold (Section~\ref{sec:threshold}) and a review of
whether the shift reflects an organizational change that belongs in the context
store.

\section{Unsupervised Anomaly Scoring}
\label{sec:unsup}

\subsection{Model}

The unsupervised model learns from historical network data without labels and
without human supervision: it is given unlabeled historical activity and allowed
to discover patterns without explicit guidance \cite[\P0028]{patent}, and it is
trained on the historical network activity data \cite[claim~6]{patent}. In some
examples it is implemented as an isolation forest \cite[\P0028]{patent}.
Alternatives include clustering, association-rule and dimensionality-reduction
models; a clustering model prepared with pre-training can group network activity
into categories, with the outliers of those categories representing anomalous
activity \cite[\P0076]{patent}.

An isolation forest \cite{iforest} builds an ensemble of random trees, each
grown on a subsample of $n$ events by repeatedly choosing a random feature and a
random split value. Anomalies are few and different, so they are isolated in
fewer splits. With $h(\mathbf{x})$ the path length of a point in a tree, the
anomaly score is
\begin{equation}
s(\mathbf{x}, n) = 2^{-\mathbb{E}[h(\mathbf{x})]/c(n)}, \qquad
c(n) = 2H(n-1) - \frac{2(n-1)}{n},
\end{equation}
where $H(i)$ is the harmonic number and $c(n)$ normalizes by the average path
length of an unsuccessful search in a binary search tree. Scores near 1
indicate anomalies, and scores well below 0.5 indicate normal points
\cite{iforest}. We write $s_U(e)$ for the score of event $e$.

\subsection{Scoring new activity}

When a new authentication request arrives, the collector passes information
about it to the unsupervised model, which evaluates it in the context of the
historical login data, and possibly the historical network data, and generates an
anomaly score \cite[\P0048]{patent}. The score describes the extent to which the
request and related data are considered anomalous \cite[\P0112]{patent}. A score
below a threshold $\tau$ suggests that the request does not represent a threat;
a score above it suggests that the request, and related activity, could be part
of a threat \cite[\P0049]{patent}. Applied this way, the model determines
whether a user's login activity has deviated from the user's baseline or from
the baseline compared with other users, and identifies anomalous failed logins,
anomalous user access to services, anomalous user agents, anomalous IP locations
and anomalous devices using the user's credentials \cite[\P0096]{patent}.

\subsection{Threshold selection}
\label{sec:threshold}

The publication compares the score with a threshold $\tau$ but does not say how
$\tau$ is chosen \cite[\P0049]{patent}. Because labeled attacks are rare, we
select the threshold from the capacity of the team that reviews alerts. Let $k$
be the number of alerts the team can review per day and $N$ the number of units
scored per day. On a recent calibration window, set
\begin{equation}
\tau = Q_{1-k/N}\bigl(\{s_U(e)\}\bigr),
\label{eq:tau}
\end{equation}
the $(1-k/N)$ quantile of the scores, so that the expected alert volume matches
the review capacity. Event units and source-window units have separate
thresholds, each with its own share of the budget. When some labels are
available, for example confirmed incidents from earlier review, $\tau$ can
instead be chosen to maximize precision at the alert budget on those labels. We
use two levels: $\tau_{\mathrm{rev}}$ sends a unit for review, and a higher
$\tau_{\mathrm{act}}$ allows automated action for categories whose actions do not
restrict access, consistent with the publication's option of requiring human
approval before an action \cite[\P0052]{patent}. Thresholds are recalibrated on
the same schedule as the model and whenever drift is detected
(Section~\ref{sec:update}).

\subsection{Using the score at authentication time}

The score can be consumed in two places. The analysis system uses it to decide
whether to act (Section~\ref{sec:response}). The service that receives the
request may also have access to the score and consider it when deciding whether
to approve the request: where the score exceeds a threshold, the service may be
less likely to authorize the request, and where it is below the threshold, more
likely \cite[\P0050]{patent}. A service may also ignore the score and decide on
its own \cite[\P0051]{patent}. This gives organizations a gradual path to
adoption: the score can begin as a signal for analysts and later become an input
to risk-based authentication, for example by requiring a stronger
authentication method for high-scoring requests.

\section{From Outliers to Threat Categories}
\label{sec:sme}

An anomaly score says that something is unusual. It does not say what. The
second stage of the framework turns outliers into named threat categories with
the help of subject matter experts.

\subsection{Expert review and tagging rules}

Over time, the analysis system identifies patterns in the anomalies the
unsupervised model finds \cite[\P0054]{patent}. An SME, working through the SME
system, requests information from the analysis system and receives historical
login data, historical network data and the anomaly scores associated with them.
The SME system evaluates that information and determines tagging rules that help
identify what type of attack is associated with the instances; it generates
labeling data for attack types and rules for how data can be labeled, based in
part on input from the experts, and returns them to the analysis system, which
uses them to associate anomaly scores with known patterns of network activity
\cite[\P0054]{patent}. In practice the expert's task is to develop an
understanding of the nature of the threat, map the anomalous traffic to an
identifiable threat in the environment, and write a rule that identifies or
labels similar activity when it occurs again \cite[\P0017]{patent}.

The analysis system can also take the first step itself. The model development
module can generate candidate rules and annotations from the outliers of the
unsupervised model, send them to the devices operated by SMEs, and receive their
modifications before any supervised model is developed from them
\cite[\P0061, \P0080]{patent}. SME feedback, both on the anomaly scores and on
the classifications, is kept in the SME data store \cite[\P0080]{patent}.

Formally, a tagging rule is a predicate $\phi_c$ over the features of an event,
or of a set of related events, that assigns category $c \in \mathcal{C}$ when it
is true. Applying the rules to an anomalous event yields $\hat{c}(e)$, or no
category if no rule fires. Figure~\ref{fig:loop} and Algorithm~1 summarize the
labeling loop, following FIG.~4 of the publication.

\begin{figure}[t]
\centering
\begin{tikzpicture}[node distance=4.5mm]
  \node[box, text width=82mm] (s1) {(402) Provide network activity, with history such as\\the prior 14 days, to the unsupervised model};
  \node[box, below=of s1, text width=82mm] (s2) {(404) Receive anomaly scores and outlier classifications};
  \node[box, below=of s2, text width=82mm] (s3) {(406) Provide the classifications to SMEs};
  \node[box, below=of s3, text width=82mm] (s4) {(408) Receive SME feedback and threat categories};
  \node[box, below=of s4, text width=82mm] (s5) {(410) Generate tagging rules and annotations,\\using organizational context};
  \node[box, below=of s5, text width=82mm] (s6) {(412) Train the supervised model};
  \node[box, below=of s6, text width=82mm] (s7) {(414) Apply both models to new network activity};
  \foreach \a/\b in {s1/s2,s2/s3,s3/s4,s4/s5,s5/s6,s6/s7} \draw[arr] (\a) -- (\b);
  \draw[arr] (s7.east) -- ++(9mm,0) coordinate (lp) |- (s1.east);
  \node[lbl, rotate=90] at ($(lp)!0.5!(lp |- s1.east)$) {new outliers for review};
\end{tikzpicture}
\caption{The expert labeling loop, after FIG.~4 of the publication
\cite[\P0099--\P0106]{patent}. Outliers that no rule or class covers feed the
next round of review.}
\label{fig:loop}
\end{figure}
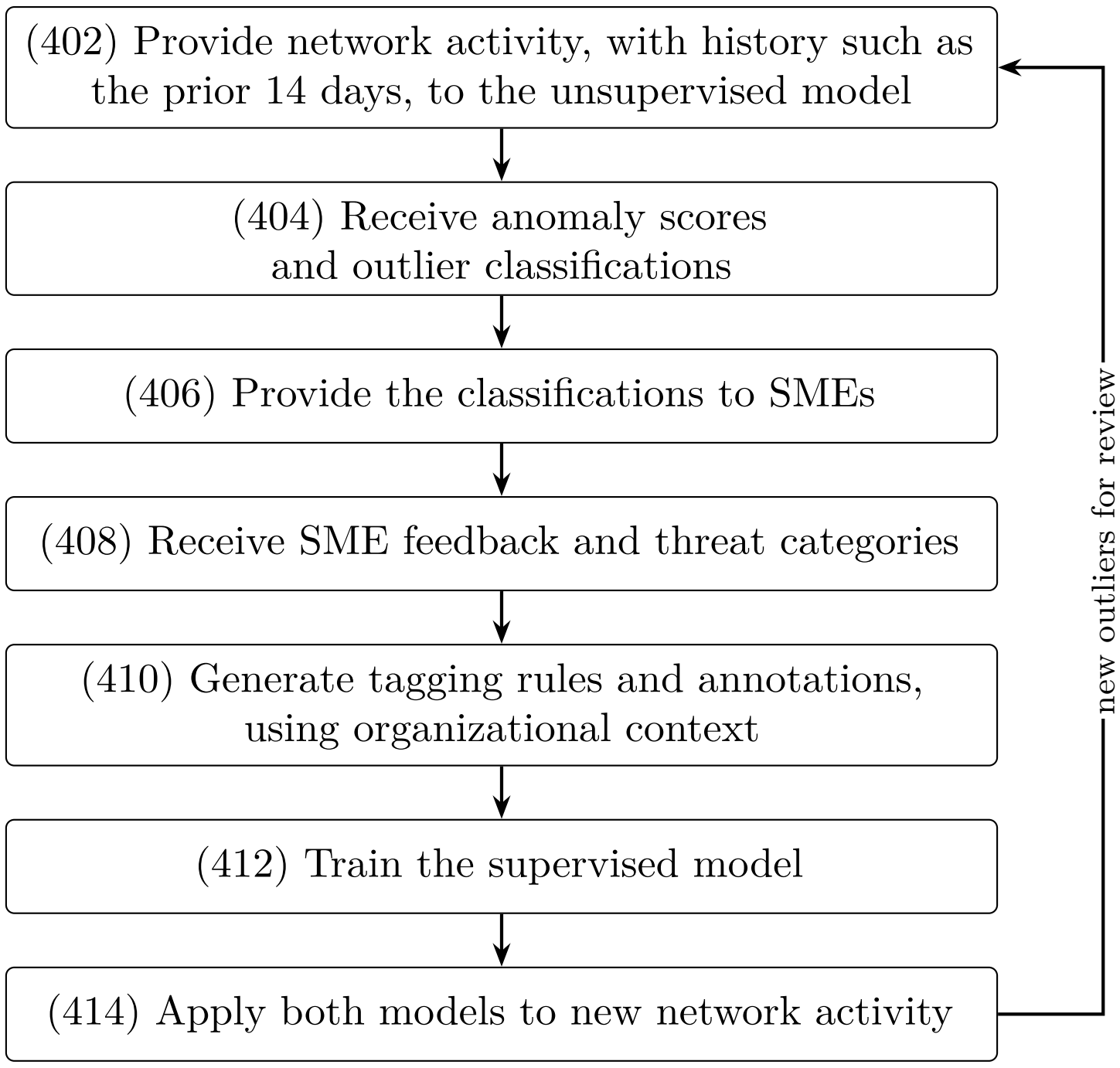

\paragraph{Algorithm 1: Building tagging rules and the supervised model.}
\begin{enumerate}[leftmargin=*,itemsep=1pt]
\item Provide network activity data from the collector data store to the
unsupervised model, together with historical data for a prior period such as
the prior 14 days \cite[\P0100]{patent}.
\item Receive anomaly scores, or outliers relative to classifications of normal
activity, identifying instances of anomalous activity \cite[\P0101]{patent}.
\item Provide the classifications to one or more SMEs to elicit feedback and
support rule development \cite[\P0102]{patent}.
\item Receive the SMEs' feedback, including changes to the classifications and
threat categories derived from them \cite[\P0102]{patent}.
\item Generate tagging rules and annotations from the feedback, applying
annotations to a set of activity data and using organizational context from the
context data store \cite[\P0103]{patent}.
\item Train a supervised model on the tagging rules and the annotated data
\cite[\P0104]{patent}.
\item Provide new activity data to both models: the unsupervised model decides
whether it is anomalous, and the supervised model attempts to classify it
\cite[\P0105]{patent}.
\end{enumerate}

\subsection{Threat categories}

Table~\ref{tab:taxonomy} lists the threat categories that the publication
names in its risk identification stage \cite[\P0097]{patent}. The ``typical
signals'' column is our illustration of how tagging rules for each category
might be written with the features of Section~\ref{sec:baseline}. It is not
taken from the publication, and an organization's experts would write rules that
fit its own environment.

\begin{table}[t]
\centering
\caption{Threat categories from the publication's risk identification stage
\cite[\P0097]{patent}, with illustrative signals a tagging rule might use.}
\label{tab:taxonomy}
\small
\begin{tabular}{@{}>{\raggedright}p{0.19\linewidth}>{\raggedright}p{0.32\linewidth}>{\raggedright\arraybackslash}p{0.41\linewidth}@{}}
\toprule
Category & Description & Typical signals (illustrative) \\
\midrule
Brute force & Brute force attacks on services or nodes & Burst of failed logins against one account or service; success rate far below the user's baseline \\
Credential checking & Scanning for compromised credentials or attempting to use them & One source trying many accounts; many distinct users from one IP address or user agent, mostly failing \\
Reconnaissance & Reconnaissance of services or nodes & One device authenticating to many more services than its baseline or its peers; new request types \\
Low-and-slow access & Unauthorized access through attacks designed to avoid detection systems & Small, persistent deviations across several attributes that no single threshold catches \\
Misconfiguration & Misconfigurations in the network & Repeated, regular failures from an automated service or device with an otherwise stable profile \\
\bottomrule
\end{tabular}
\end{table}

\subsection{Discovering new categories}

The categories are not a closed list. The system uses predetermined categories,
based on known types of malicious activity and public databases of attack
vectors and techniques, together with categories that the models identify or
generate; in particular, it may determine new categories when the unsupervised
model identifies a novel attack vector \cite[\P0059]{patent}. The model
development module correlates the unsupervised model's classifications with
rationales for the anomalous activity, such as an attack or misconfigured
network resources, and uses SME feedback to identify the categories
\cite[\P0083]{patent}. Mapping categories to a public taxonomy such as MITRE
ATT\&CK \cite{attack} is a natural way to seed the predetermined set and to
communicate detections to other security tools.

\section{Supervised Classification in Tandem}
\label{sec:tandem}

\subsection{Training}

To the extent that the tagging rules label known patterns of anomalous behavior,
the labeled patterns can train a supervised model to predict those patterns from
network activity \cite[\P0017]{patent}. The analysis system applies the tagging
rules to a set of historical network activity data and trains the supervised
model on the result, so that it can identify and classify anomalous activity into
a threat category \cite[\P0062]{patent}. Training data can include the
annotated data, the rules themselves and organizational information and
context \cite[\P0081]{patent}. Once trained, the supervised model does not
replace the unsupervised model; it operates in conjunction with it and
supplements its predictions \cite[\P0055]{patent}.

\subsection{Reject option}
\label{sec:reject}

The publication requires that activity which the supervised model cannot assign
to a category is still acted on \cite[\P0057, \P0084]{patent}, but it does not
say how the model decides that it cannot assign a category. A classifier trained
on $|\mathcal{C}|$ categories will otherwise assign every input to one of them,
including inputs from a category it has never seen. We specify a reject option
with two tests.

\paragraph{Training set.} The classifier is trained only on units that the
unsupervised model flagged, labeled by the tagging rules or by expert
annotation. Units on which two or more rules assign different categories are
sent to experts rather than used for training; where rules overlap heavily, a
label model of the kind used in weak supervision \cite{snorkel} can combine them
instead. Alerts that experts dismissed as benign form an additional class,
\emph{expected}, so that the classifier also learns to recognize recurring false
positives. Any probabilistic multi-class classifier can be used; gradient-boosted
trees suit the tabular features of Section~\ref{sec:features}.

\paragraph{Confidence test.} Let $p(c \mid \mathbf{x})$ be the calibrated class
probabilities. Following Chow's rule \cite{chow}, the prediction
$\hat{c} = \arg\max_c p(c \mid \mathbf{x})$ passes the confidence test if
$p(\hat{c} \mid \mathbf{x}) \ge \theta_{\hat{c}}$.

\paragraph{Distance test.} A confident output does not guarantee that the input
resembles the training data \cite{hendrycks}. The prediction passes the distance
test if the distance from $\mathbf{x}$ to the training units of category
$\hat{c}$ is within the range seen in training. We use the mean distance to the
$k$ nearest training units of $\hat{c}$ and accept if it is at most
$\delta_{\hat{c}}$, the 95th percentile of the same quantity computed on held-out
training units of $\hat{c}$.

\paragraph{Decision.} The classifier accepts $\hat{c}$ only if both tests pass;
otherwise it rejects, and we write $\hat{c}(e)=\bot$. The thresholds
$\theta_c$ are chosen on held-out labeled units so that the precision of accepted
predictions in each category meets a target set by the security team, for
example 0.9. The leave-one-category-out test of Section~\ref{sec:eval} measures
how often units from an unseen category are rejected, which is the property the
tandem design depends on.

\subsection{Inference}

At inference time the collector sends information about each new request to
both models \cite[\P0056]{patent}. The unsupervised model decides whether the
request is anomalous. The supervised model decides whether it matches a known
threat or attack pattern and, if so, outputs a classification identifying the
pattern, on which the analysis system acts \cite[\P0056]{patent}. When the
unsupervised model also finds the request anomalous, the action taken in
response to the anomaly may be subsumed by the action taken in response to the
classification \cite[\P0056]{patent}. When a request is a threat but matches no
known pattern, the tagging rules and the supervised model cannot identify it,
but the unsupervised model's score can still indicate that it is anomalous, and
the analysis system acts on that score \cite[\P0057]{patent}.
Algorithm~2 and Table~\ref{tab:tandem} state this logic.

\paragraph{Algorithm 2: Tandem inference for a new unit $e$.}
\begin{enumerate}[leftmargin=*,itemsep=1pt]
\item Compute the anomaly score $s_U(e)$ with the unsupervised model. If
$s_U(e) \le \tau_{\mathrm{rev}}$ and no tagging rule fires, take no action
\cite[\P0112]{patent}.
\item Otherwise apply the tagging rules and the supervised classifier with its
reject option \cite[\P0113]{patent}, giving $\hat{c}(e) \in \mathcal{C} \cup
\{\textit{expected}, \bot\}$.
\item If $\hat{c}(e) \in \mathcal{C}$, take the action for category $\hat{c}(e)$;
any action for the anomaly alone is subsumed \cite[\P0056]{patent}.
\item If $\hat{c}(e) = \textit{expected}$, record the unit without action and
include it in periodic reporting.
\item If $\hat{c}(e) = \bot$, treat $e$ as an unclassified anomaly: take the
default mitigating action \cite[claim~3]{patent}, which requires approval unless
$s_U(e) > \tau_{\mathrm{act}}$, and queue $e$ for expert review as a candidate
for a new category.
\end{enumerate}

\begin{table}[t]
\centering
\caption{Outcome of tandem inference.}
\label{tab:tandem}
\small
\begin{tabular}{@{}>{\raggedright}p{0.22\linewidth}>{\raggedright}p{0.22\linewidth}>{\raggedright\arraybackslash}p{0.46\linewidth}@{}}
\toprule
Supervised model & Unsupervised model & Outcome \\
\midrule
Known category & Anomalous & Act for the category; the anomaly action is subsumed \\
Known category & Not anomalous & Act for the category (rules applied to all units) \\
Expected & Anomalous & Record without action; include in reporting \\
Rejected ($\bot$) & Anomalous & Act on the anomaly, with approval unless the score is very high; send for review as a possible new category \\
No category & Not anomalous & No action \\
\bottomrule
\end{tabular}
\end{table}

In our instantiation the tagging rules are applied to every unit, but the
classifier is applied only to units that are flagged or on which a rule fires,
because it is trained on flagged units and its outputs on ordinary traffic would
be out of distribution. The publication applies both models to each request
\cite[\P0056]{patent}; the two coincide when the rules cover every known
category.

An unclassified anomaly is recognized in one of two ways: the rules are applied
and none of them classifies the activity \cite[claim~4]{patent}, or the
supervised model cannot assign it to any category \cite[\P0084]{patent}, which
in our instantiation means that it rejects. Either way the activity is still
acted on. This is what allows the design, in principle, to combine the
consistency of rules for known categories with a response to activity outside
them \cite[\P0057, \P0106]{patent}. Two limits apply. An attack that does not
look anomalous in the chosen features is missed by both paths, and an attack
from a new category that the classifier confidently assigns to a known one is
mislabeled; the reject option reduces but does not remove this risk. We make no
claim about how often either happens; Section~\ref{sec:eval} describes how to
measure it.

\subsection{Iterating the loop}

The loop of Figure~\ref{fig:loop} is designed to be repeated. Unclassified
anomalies reach experts, who may write new rules; new rules label more data; the
supervised model is retrained with more categories. The publication anticipates
that, as more techniques are analyzed and labeled, the framework can translate
more kinds of anomalous behavior into specific attack techniques with less
supervision \cite[\P0017, \P0019]{patent}. This is an expectation about the
design, not a demonstrated property. Whether repeated rounds improve
categorization depends on how much expert time is available, on the quality of
the rules, and on whether errors in the rules propagate into the classifier; a
loop of this kind can also reinforce its own mistakes if dismissed alerts are
mislabeled. We treat the number of rounds, the expert effort per round and the
categorization quality after each round as quantities to measure
(Section~\ref{sec:eval}).

\section{Response: Remediation and Reporting}
\label{sec:response}

\subsection{Choosing an action}

When the score or the classification indicates a potential threat, the analysis
system sends a control signal to the remediation system, which takes an action
chosen to counter the threat or to limit the harmful effects of an unknown one
\cite[\P0052]{patent}. The action depends on the threat category
\cite[\P0060]{patent}. Table~\ref{tab:actions} lists the classes of action named
in the risk remediation stage of the publication \cite[\P0098]{patent}.

\begin{table}[t]
\centering
\caption{Classes of remediation action from the publication's risk
remediation stage \cite[\P0098]{patent}.}
\label{tab:actions}
\small
\begin{tabular}{@{}>{\raggedright}p{0.32\linewidth}>{\raggedright\arraybackslash}p{0.60\linewidth}@{}}
\toprule
Action & Purpose \\
\midrule
Security operations & Execute security operations in response to the threat \\
Remediate security incidents & Resolve the incident using one or more techniques \\
Identity and access control & Identify and act on access control for network components \\
Access policy remediation & Correct the access policies of services and nodes \\
Infrastructure management & Manage the infrastructure of the network \\
Configuration remediation & Modify the configuration of network components \\
\bottomrule
\end{tabular}
\end{table}

Actions can be directed at the network or at the source of the request.
Network-side actions include adjusting configurations of the network and the
availability of its resources, nodes and services \cite[\P0052]{patent}, and
activating security components or remediating the configuration of services and
nodes \cite[\P0060]{patent}. When activity in a category is consistent with a
number of attackers targeting one service, for example, the remediation module
may apply an updated security patch to harden that service
\cite[\P0085]{patent}. Source-side actions include restricting the source
device's access to services and nodes and revoking any authentication already
granted to it \cite[\P0052]{patent}. A less obvious source-side option is to
keep watching: the remediation system may continue to monitor a device even when
its score indicates a threat, to learn more about how it operates and what its
motives are \cite[\P0052]{patent}.

\subsection{Controlling downstream systems}

The analysis system acts by sending control signals to other systems. It may
instruct the remediation system to perform a specific operation, or send a series
of signals to the remediation system, to controlled systems downstream, or to
both, which determine that the signals contain instructions to mitigate the
threat \cite[\P0053]{patent}. A controlled system may be instructed to modify the
configuration of the network to mitigate the threat \cite[claim~11]{patent}. In
this way the analysis system controls the systems that act, based on the anomaly
score and the classified category \cite[\P0114]{patent}.

\subsection{Automation and human approval}

Actions may be taken automatically or only after a human administrator or
security analyst approves them \cite[\P0052]{patent}, and steps of the method
may likewise be performed automatically or in response to input or another event
\cite[\P0115]{patent}. We regard this as an essential deployment control. A false
positive that blocks a legitimate user from a critical application has a real
cost, so organizations will typically automate actions for well-understood,
high-confidence categories and require approval for unclassified anomalies and
for actions that restrict access.

\subsection{Reporting}

The reporting module produces reports and alerts on a periodic schedule, in
response to classification into particular categories, or both
\cite[\P0086]{patent}. For a severe category it sends an immediate alert rather
than waiting for the periodic report \cite[\P0086]{patent}; when the models
classify an event as a relatively serious security deficiency, the alert can go
to multiple devices operated by the network's cybersecurity administrators
\cite[\P0060]{patent}. For example, when recent activity is classified into a
category representing a distributed denial of service attack against a service,
the reporting module generates a report about the attack and sends it to
recipient devices \cite[\P0086]{patent}. Reports also support the
organization's compliance reporting through the SIEM system
\cite[\P0031]{patent}.

\section{End-to-End Operation}
\label{sec:e2e}

Figure~\ref{fig:pipeline} shows the seven-stage pipeline of FIG.~3 of the
publication, and Figure~\ref{fig:flow} the operations of FIG.~5. We walk through
them with three requests, following the publication's examples.

\begin{figure}[t]
\centering
\begin{minipage}[t]{0.48\linewidth}
\centering
\begin{tikzpicture}[node distance=3.5mm]
  \node[box, minimum width=66mm] (p1) {1. User accessing suite of applications};
  \node[box, below=of p1, minimum width=66mm] (p2) {2. Analyze logon attributes};
  \node[box, below=of p2, minimum width=66mm] (p3) {3. Baseline user activity, last 14 days};
  \node[box, below=of p3, minimum width=66mm] (p4) {4. Compare user activity to other users};
  \node[box, below=of p4, minimum width=66mm] (p5) {5. Logon anomaly detection model};
  \node[box, below=of p5, minimum width=66mm] (p6) {6. Risk identification};
  \node[box, below=of p6, minimum width=66mm] (p7) {7. Risk remediation};
  \foreach \a/\b in {p1/p2,p2/p3,p3/p4,p4/p5,p5/p6,p6/p7} \draw[arr] (\a) -- (\b);
\end{tikzpicture}
\caption{The seven-stage pipeline, after FIG.~3 of the publication
\cite[\P0090--\P0098]{patent}.}
\label{fig:pipeline}
\end{minipage}\hfill
\begin{minipage}[t]{0.48\linewidth}
\centering
\begin{tikzpicture}[node distance=3.5mm and 4mm]
  \node[box, minimum width=50mm] (f1) {(501) Obtain historical activity};
  \node[box, below=of f1, minimum width=50mm] (f2) {(502) Determine the baseline};
  \node[box, below=of f2, minimum width=50mm] (f3) {(503) Collect new activity};
  \node[box, below=of f3, minimum width=50mm] (f4) {(504) Anomalous?};
  \node[box, right=4mm of f4, font=\scriptsize] (no) {No action};
  \node[box, below=6mm of f4, minimum width=50mm] (f5) {(505) Classify into a category};
  \node[box, below=of f5, minimum width=50mm] (f6) {(506) Act to mitigate the threat};
  \foreach \a/\b in {f1/f2,f2/f3,f3/f4,f5/f6} \draw[arr] (\a) -- (\b);
  \draw[arr] (f4) -- node[lbl, right=2pt]{yes} (f5);
  \draw[arr] (f4) -- node[lbl, above=2pt]{no} (no);
\end{tikzpicture}
\caption{Operations of the analysis system, after FIG.~5 of the publication
\cite[\P0109--\P0114]{patent}.}
\label{fig:flow}
\end{minipage}
\end{figure}

\paragraph{Building the baseline.} Services report the authentication requests
they process as login data, and nodes report network data; the analysis system
stores both (step 501) \cite[\P0109]{patent}. It determines normal behavior for
each user and device, for each type of user, such as marketing or sales
personnel, and for each type of device, and uses this as a baseline in various
contexts (step 502) \cite[\P0110]{patent}.

\paragraph{Request A: a known threat.} A device sends a request to a service.
The analysis system receives information about it from the service, or observes
it on the network directly (step 503) \cite[\P0111]{patent}, and compares it and
related activity with the baseline for that user and device. The unsupervised
model finds it anomalous and outputs a score describing how anomalous it is
(step 504) \cite[\P0112]{patent}. Tagging rules translate the score into a known
category, or the supervised model classifies the request directly (step 505)
\cite[\P0113]{patent}, and the analysis system sends control signals to the
remediation system, which may in turn control downstream systems (step 506)
\cite[\P0114]{patent}.

\paragraph{Request B: a known pattern, after training.} Once the supervised model
is trained, a second request is scored by both models. The supervised model
recognizes a known pattern of threatening activity and outputs its
classification, and the analysis system acts on it; the action for the anomaly
is subsumed \cite[\P0056]{patent}.

\paragraph{Request C: an unknown threat.} A third request is a threat but matches
no pattern the experts have identified. Neither the rules nor the supervised
model can name it. If the unsupervised model's score indicates that it is
anomalous, the analysis system acts on that score \cite[\P0057]{patent}; if the
request does not look anomalous in the chosen features, neither path catches it.
This request then becomes material for the next round of expert review.

\section{Illustrative Scenarios}
\label{sec:scenarios}

The scenarios below are illustrations, not results; Section~\ref{sec:poc}
reports the proof of concept. They show how the
framework treats the kinds of activity the publication describes, and which
parts of the method each relies on.

\begin{enumerate}[leftmargin=*,itemsep=3pt]
\item \textbf{Brute force against one service.} A burst of failed logins
against one account drives its success rate far below its baseline and its
failed-login count far above its peers. The event scores as anomalous, a
brute-force rule fires and the supervised model names it, so the alert reaches
the SIEM with its category and the configured response, such as restricting the
source, runs \cite[\P0052, \P0056]{patent}.
\item \textbf{Credential checking across accounts.} One source tries a list of
accounts, each only a few times. No single account looks unusual, but the
source's count of distinct users and its failure rate are extreme relative to
other sources, and a credential-checking rule written over those features
classifies it \cite[\P0097]{patent}.
\item \textbf{Low-and-slow access.} An account makes occasional attempts that
stay under the levels that trigger typical security systems
\cite[\P0005]{patent}. Each attribute deviates only slightly, but because many
attributes are compared with the account's own baseline, the combination can
stand out even when no single attribute crosses a threshold
\cite[\P0043]{patent}. Whether it does so in practice, and at what alert volume,
is an empirical question.
\item \textbf{Probing after access.} A device that signed into one service starts
trying several others in a way outside the norm. This may indicate
reconnaissance, and it also shows which services interest the device and which
may be vulnerable \cite[\P0043]{patent}. Remediation can restrict the device, or
keep monitoring it to learn more \cite[\P0052]{patent}.
\item \textbf{Misconfiguration.} An automated service begins failing to
authenticate on a regular schedule after a change. The pattern is anomalous but
not malicious; classifying it as misconfiguration routes it to configuration
remediation rather than to incident response \cite[\P0097, \P0098]{patent}.
\item \textbf{Expected business activity.} A finance team's quarter-end activity
spans unusual applications at unusual hours. Organizational context marks it as
known, so it is filtered out before it becomes an alert \cite[\P0019]{patent}.
\item \textbf{A previously unseen technique.} Activity that the unsupervised
model scores above threshold, and that no rule covers and the classifier
rejects, is still acted on, and expert review of it can produce a new category
\cite[\P0018, \P0059, \P0084]{patent}. Activity of this kind that scores below
threshold, or that the classifier confidently assigns to a known category, is
not handled this way.
\end{enumerate}

\section{Evaluation Protocol}
\label{sec:eval}

We set out how the framework should be evaluated so that implementations can be
compared on equal terms. Section~\ref{sec:poc} applies a reduced form of this
protocol.

\paragraph{Data.} Evaluation requires authentication events with known
malicious activity. The Los Alamos comprehensive multi-source data set
\cite{lanl} covers 58 consecutive days of de-identified events from an
enterprise network, including authentication events and a small set of
authentication events marked as red-team compromise activity. It suits the
per-user baselining of Section~\ref{sec:baseline}: the first 14 days form the
initial baseline window, later days are scored with the rolling procedure of
Section~\ref{sec:update}, and source computers define the source windows of
Section~\ref{sec:features}. Its red-team events carry no threat category, so
categorization requires a second source. Synthetic data, generated by injecting
each threat category of Table~\ref{tab:taxonomy} into the benign traffic at
controlled rates, gives a label for every category, including low-and-slow
patterns that are rare in real data. Tagging rules should be written by a person
who sees only the training period, so that rule writing does not leak test
labels.

\paragraph{Tasks and metrics.} We propose four tasks.
\begin{enumerate}[leftmargin=*,itemsep=1pt]
\item \emph{Detection}: rank events by $s_U(e)$, and report the area under the
ROC and precision--recall curves and the precision at the number of alerts an
analyst team can review per day.
\item \emph{Categorization}: for detected malicious events, report the
macro-averaged $F_1$ score over threat categories, for the tagging rules and for
the supervised model.
\item \emph{Novelty}: hold out one category at a time from rule writing and
training, and report the fraction of held-out events that the tandem design
still acts on as unclassified anomalies, the fraction the classifier wrongly
assigns to a known category, and, for comparison, the same fractions with the
confidence test alone and with no reject option.
\item \emph{Iteration}: simulate rounds of the loop of Figure~\ref{fig:loop}
in which rules are added for the most frequent unclassified pattern after each
round, and report categorization quality and expert effort, measured as rules
written and alerts reviewed, per round.
\end{enumerate}
Thresholds should be set by Equation~\eqref{eq:tau} for a stated alert budget,
and results reported at that budget, so that methods are compared at equal
analyst workload.
Analyst workload, measured as alerts per day and as the share of alerts that
arrive with a category, should be reported alongside accuracy, because the
framework's purpose is to make alerts actionable as well as accurate.

\paragraph{Baselines and ablations.} Natural baselines are static threshold
rules, the unsupervised model alone and the supervised model alone. Ablations
should remove peer comparison, remove organizational context, and vary the
baseline window across a week, 14 days and a month \cite[\P0094]{patent}, to
show how much each element contributes.

\section{Proof of Concept on LANL Data}
\label{sec:poc}

We ran a modest proof of concept that follows the protocol of
Section~\ref{sec:eval} where the data allow it. Its purpose is to check that the
reference method can be implemented end to end on public data, and to report
honestly how it compares with simple baselines. It is not a tuned system, and its
results do not show that the framework is effective. The code, which runs in a
few minutes on a laptop, is available from the authors on request.

\subsection{Setup}

\paragraph{Data.} We used the authentication and red-team files of the Los
Alamos data set \cite{lanl}, restricted to the first 30 days, which contain the
red-team activity. We kept logon events of user accounts (names beginning with
``U''), dropped computer accounts and log-off events, and kept every account that
appears in the red-team file together with a fixed 10\% hash sample of the other
accounts. This leaves 16{,}081{,}176 events from 1{,}231 users, 98 of whom have
red-team events. Keeping every red-team account raises the share of compromised
users far above its share in the full population, so absolute precision figures
here are not those of a deployment.

\paragraph{Units and features.} The unit of analysis is the user-day, not the
single event of Section~\ref{sec:features}, because LANL events carry no user
agent, application or request type, and because red-team labels are sparse at the
event level. A user-day is positive when the red-team file lists that user on that
day. For each user-day we computed four groups of features:
\emph{counts} (events, failures, failure rate, distinct source and destination
computers, distinct authentication and logon types, active hours, the coefficient
of variation of inter-arrival times and the peak events per hour);
\emph{novelty and rarity} against the user's own window of $W$ days (new
destinations, sources, authentication and logon types, and the rarity of the
destinations, sources and hours used);
\emph{ratios} of the day's counts to the user's baseline mean, shrunk toward the
peer-group mean as in Equation~\eqref{eq:shrink} with $\kappa=3$;
and \emph{peer} scores, the robust $z$-score of Equation~\eqref{eq:peer} against
same-day members of the user's peer group. LANL has no roles or organizational
structure, so peer groups are eight $k$-means clusters of each user's activity
profile over the first $W$ days, and organizational context
(Section~\ref{sec:context}) is not used.

\paragraph{Detection.} An isolation forest \cite{iforest} with 200 trees is
trained on the user-days of the previous $W=14$ days, excluding user-days flagged
earlier (quarantine, Section~\ref{sec:update}), and retrained every 7 days. Days
14 to 29 are scored. The review threshold is set by Equation~\eqref{eq:tau} for a
budget of 10 alerts per day. The test period holds 13{,}102 user-days, of which
60 (0.46\%), from 47 users, are red-team positives.

\paragraph{Categorization.} The red-team events carry no threat category, so we
injected synthetic activity of five categories into real benign user-days with at
least five days of history: brute force (40 to 150 network logons to one usual
destination within an hour, nearly all failing), credential use (a few successful
logons from one or two previously unseen source computers), reconnaissance (15 to
40 logons to previously unseen destinations within two hours), low-and-slow
access (two or three previously unseen destinations, one or two logons each,
at night) and misconfiguration (50 to 150 regularly spaced failing service
logons with an authentication type the user has not used). Days 14 to 21 form the
training split and days 22 to 29 the test split, with 40 injections per category
in each. Tagging rules have a fixed structure per category, one to four threshold
conditions on the features above, and each threshold is set from the 10th or
90th percentile of that category's training injections, as an analyst would after
reviewing them. The supervised model is a gradient-boosted tree classifier with
the two-part reject option of Section~\ref{sec:reject}: a per-class confidence
threshold chosen for 0.9 precision on a validation split, and a per-class
nearest-neighbor distance test at the 95th percentile of validation distances.

\subsection{Detection of red-team activity}

Table~\ref{tab:pocA} compares the framework's detector with two baselines, a rule
that ranks user-days by the number of new destinations and an isolation forest
on the count features alone, and with ablations.

\begin{table}[t]
\centering
\small
\caption{Detection of red-team user-days on LANL, days 14 to 29 (13{,}102
user-days, 60 positive from 47 users). P@10 and R@10 are the precision and recall
of the ten highest-scoring user-days per day; ``users'' counts red-team users
found in those ten. The last three columns use the threshold of
Equation~\eqref{eq:tau} with a budget of 10 alerts per day. The average precision
of a random ranking is 0.005.}
\label{tab:pocA}
\begin{tabular}{@{}>{\raggedright}p{0.28\linewidth}cccccccc@{}}
\toprule
Method & AUC & AP & P@10 & R@10 & Users & Alerts/day & Prec. & Rec.\\
\midrule
Rule: new destinations & 0.761 & 0.015 & 0.019 & 0.050 & 3 & -- & -- & --\\
Isolation forest, counts only & 0.764 & 0.024 & 0.069 & 0.183 & 5 & 15.8 & 0.047 & 0.200\\
Framework (all features) & 0.734 & 0.022 & 0.044 & 0.117 & 2 & 9.5 & 0.046 & 0.117\\
\quad without peer scores & 0.769 & 0.021 & 0.031 & 0.083 & 2 & 8.1 & 0.031 & 0.067\\
\quad without novelty and rarity & 0.729 & 0.028 & 0.069 & 0.183 & 4 & 13.1 & 0.052 & 0.183\\
\quad window $W=7$ days & 0.775 & 0.021 & 0.031 & 0.083 & 2 & 16.3 & 0.038 & 0.167\\
\quad window $W=21$ days & 0.696 & 0.023 & 0.056 & 0.150 & 3 & 5.9 & 0.074 & 0.117\\
\bottomrule
\end{tabular}
\end{table}

The result is negative for the framework's detector. With all features it scores
below both baselines on ROC-AUC and on recall in the top ten alerts per day, and
removing the novelty and rarity features does not hurt it. Every method ranks
red-team days above chance, with an average precision three to six times that of a
random ranking, but almost every alert is a false alarm. Two of the
framework's elements do not translate well to LANL. The peer groups are
clusters, not the roles and organizational units the framework assumes, and the
user-day unit averages a few red-team logons into a day of ordinary activity.
We did not tune features or hyperparameters on the test period, and we report
the configuration as specified.

\subsection{Categorization and the reject option}

\paragraph{End to end.} We first ran the categorization stage on the injected
units that the detector flagged under its budget. This setting turned out to be
uninformative. Injections averaged 25 per day against a budget of 10 alerts per
day, so most could not be flagged: the share flagged was 52.5\% for
misconfiguration, 20\% for reconnaissance, 19\% for brute force and 1\% for
credential use and low-and-slow access. On the 56 flagged test injections the
tagging rules reached a macro-$F_1$ of 0.44. The classifier was trained on only
three or four flagged examples per category and either labeled every test unit
as expected or rejected every one, a macro-$F_1$ of zero. We report this as
inconclusive: it reflects the design of the experiment, not the method.

\paragraph{Categorization given detection.} We then evaluated the categorization
stage on its own, as if the detector had surfaced every injected unit. Rules and
classifier are trained on all 200 injections of the training split, the
``expected'' class is formed from the ten highest-scoring benign user-days of each
day, which stand in for alerts an analyst would dismiss, and testing uses all 200
test injections and the corresponding benign units. This changes the evaluation,
not the method. Table~\ref{tab:pocB} reports the results by category.

\begin{table}[t]
\centering
\small
\caption{Categorization of synthetic attacks injected into LANL user-days, given
detection (40 test injections per category). ``Score AUC'' measures how well the
anomaly score separates each category from benign user-days. The last three
columns give per-category recall.}
\label{tab:pocB}
\begin{tabular}{@{}>{\raggedright}p{0.25\linewidth}cccc@{}}
\toprule
Category & Score AUC & Rules & Classifier & Classifier with reject\\
\midrule
Brute force & 0.969 & 0.500 & 0.800 & 0.725\\
Credential use & 0.700 & 0.775 & 0.925 & 0.925\\
Reconnaissance & 0.966 & 0.800 & 0.975 & 0.900\\
Low-and-slow access & 0.641 & 0.650 & 0.775 & 0.775\\
Misconfiguration & 0.990 & 0.425 & 0.900 & 0.675\\
\midrule
Macro-$F_1$ & -- & 0.72 & 0.88 & 0.85\\
\bottomrule
\end{tabular}
\end{table}

The anomaly score separates the loud categories, brute force, reconnaissance and
misconfiguration, from benign activity (AUC 0.97 to 0.99) but not the quiet ones,
credential use and low-and-slow access (0.70 and 0.64). The tagging rules cover
74\% of test injections and conflict on 12\%. The classifier reaches a macro-$F_1$
of 0.88 and an accuracy of 87\% with no reject option. The confidence test alone
rejects 11\% of units and raises the accuracy of the rest to 89\%; adding the
distance test rejects 22.5\% and raises it to 91\%. The confidence thresholds of
the five attack categories came out at zero, because the classifier already
reached 0.9 validation precision on them at any confidence, so in practice the
confidence test rejects only units predicted as expected, and nearly all
rejection of attack categories comes from the distance test. Of the seven
red-team user-days in the flagged set, the classifier labeled six as expected and
rejected one.

\paragraph{Unseen categories.} Table~\ref{tab:loco} reports the novelty task:
each category in turn is removed from training, and we record what the classifier
does with its test injections.

\begin{table}[t]
\centering
\small
\caption{Leave-one-category-out test (40 test injections of the held-out
category). Each cell gives the share rejected / labeled expected / labeled as
another attack category.}
\label{tab:loco}
\begin{tabular}{@{}>{\raggedright}p{0.22\linewidth}ccc@{}}
\toprule
Held-out category & No reject & Confidence test & Confidence and distance\\
\midrule
Brute force & 0 / 0.85 / 0.15 & 0.25 / 0.60 / 0.15 & 0.30 / 0.60 / 0.10\\
Credential use & 0 / 0.30 / 0.70 & 0 / 0.30 / 0.70 & 0.30 / 0.30 / 0.40\\
Reconnaissance & 0 / 0.18 / 0.83 & 0.10 / 0.08 / 0.83 & 0.93 / 0.08 / 0\\
Low-and-slow access & 0 / 0.83 / 0.18 & 0 / 0.83 / 0.18 & 0.03 / 0.83 / 0.15\\
Misconfiguration & 0 / 0.20 / 0.80 & 0.03 / 0.18 / 0.80 & 0.20 / 0.18 / 0.63\\
\bottomrule
\end{tabular}
\end{table}

The reject option is the part of the framework designed to keep unseen attacks
from being forced into a known category. Here it works for one category and
mostly fails for the others. With both tests, 92.5\% of unseen reconnaissance is
rejected and none is mislabeled as another attack. For the other categories, 3\%
to 30\% is rejected, and much of the rest is labeled expected, including 60\% of
unseen brute force and 83\% of unseen low-and-slow access, which in a deployment
would mean these attacks are dismissed as normal.

\subsection{Limitations of the proof of concept}

Three limitations bound what these results can show. First, the attacks in the
categorization experiment are synthetic, and the same code generated them and
fixed the structure of the tagging rules, so high categorization scores are to be
expected; they show that the stage works mechanically, not that it would work on
real attacks. Second, the experiment uses user-days, clustered peer groups and no
organizational context, so it tests a reduced form of the framework. Third, it
covers one data set and one configuration. The detection configuration was
fixed before any results were seen. The categorization-given-detection test was
added after the first run showed the end-to-end setting to be uninformative, and
nothing was tuned on test results. Within those limits, the evidence is that the
framework's detector did not outperform simple baselines on real red-team
activity, that the categorization stage can work when it receives enough
examples, and that the reject option did not reliably catch unseen categories.

\section{Discussion and Limitations}
\label{sec:discussion}

\paragraph{Why tandem.} Training an unsupervised model and using it to develop
tagging rules lets an organization build a classifier for a system as complex as
a SaaS platform without the error-prone, hard-to-maintain ground-up development
of a rules-based model, and that running the two in tandem combines the
consistency of rules with the unsupervised model's sensitivity to activity the
rules do not describe; this is the argument of the publication
\cite[\P0063]{patent}. Our proof of concept does not support it for detection:
the framework's detector did not outperform a count-only isolation forest or a
simple rule (Section~\ref{sec:poc}). What the design does
guarantee by construction is narrower: an alert that receives a category can be
explained by the rule or class that assigned it, and an anomaly that receives no
category is not discarded.

\paragraph{Baseline quality.} A baseline is only as good as the window it is
built from. Users with little history, such as new employees, have weak
baselines. If an attacker is already active during the window, and the activity
was never flagged, the baseline absorbs it. The shrinkage, quarantine and
rolling recomputation of Section~\ref{sec:update} reduce these problems but do
not remove them.

\paragraph{Expert effort and drift.} Tagging rules need expert time to write and
to keep current as techniques change, and rules that are too broad will label
benign activity and propagate the error into the supervised model. Generating
candidate rules automatically and having experts edit them
\cite[\P0061, \P0080]{patent} reduces this cost, and tracking which rules fire,
and how often their labels are overturned, identifies rules that need revision.

\paragraph{Operational risk.} Learning-based detection has costs that are easy
to underestimate: attacks are rare, false positives are expensive and analysts
need explanations \cite{sommer}. Named categories, organizational context and
human approval for disruptive actions \cite[\P0052]{patent} are the framework's
answers to these costs, but thresholds still trade missed attacks against alert
volume and must be tuned for each environment.

\paragraph{Privacy.} Logon data identifies people, devices and locations.
Deployments should restrict access to the collector data store, retain data only
as long as the baseline and audit requirements need, and apply the same
governance to SME annotations as to the raw data.

\section{Conclusion}
\label{sec:conclusion}

We described a technical framework for detecting security threats from logon
data in SaaS systems. It builds baselines for users, categories, devices and contexts from
authentication attributes and node activity, compares users with their peers,
and scores new activity with an unsupervised model. Subject matter experts turn
the outliers into tagging rules and threat categories, organizational context
removes expected activity, and a supervised model trained on rule-labeled data
runs in tandem with the unsupervised one and rejects what it cannot confidently
categorize, so that anomalous activity outside the known categories is still
routed to response and review. Each outcome drives remediation, control of
downstream systems or severity-dependent reporting. Beyond the patent
publication, we specified the feature construction, threshold selection, reject
option and baseline updating, and positioned the framework against prior work.
A proof of concept on the Los Alamos data found that the framework's detector
did not outperform simple baselines on real red-team activity, that the
categorization stage worked on synthetic attacks once it had enough examples, and
that the reject option caught one unseen category well and the others poorly. We
have therefore not shown that the framework detects attacks more effectively than
existing approaches. Next steps are an evaluation on data with roles and
organizational context, event-level units in place of user-days, and attacks
labeled by category rather than synthesized.

\appendix
\section{Mapping of the Publication's Claims}
\label{app:claims}

Table~\ref{tab:claims} maps the claims of US~2026/0006036~A1 to the sections of
this paper. Claims 12--19 recite a computing system, and claim 20
computer-readable media, that perform the same operations as the method claims.

\begin{table}[h]
\centering
\caption{Claims of the publication and where this paper describes them.}
\label{tab:claims}
\small
\begin{tabular}{@{}>{\raggedright}p{0.14\linewidth}>{\raggedright}p{0.58\linewidth}>{\raggedright\arraybackslash}p{0.18\linewidth}@{}}
\toprule
Claims & Subject & Sections \\
\midrule
1, 12, 20 & Baseline from authentication traffic, unsupervised detection, classification into a threat category, mitigating action & \ref{sec:baseline}--\ref{sec:response} \\
2, 13 & SME-created rules for each threat category, applied to classify & \ref{sec:sme} \\
3, 4, 14, 15 & Anomalous activity that no rule classifies is still acted on & \ref{sec:tandem} \\
5, 16 & Tagging rules created from input at an SME computing system & \ref{sec:sme} \\
6, 17 & Unsupervised model trained on historical activity & \ref{sec:unsup} \\
7, 18 & Supervised model trained on labeled historical activity & \ref{sec:tandem} \\
8, 19 & Normal logon behavior per user and per type of user & \ref{sec:baseline} \\
9 & Login attributes relative to other users & \ref{sec:baseline} \\
10 & Organizational context in the baseline & \ref{sec:baseline} \\
11 & Control signals to a controlled system to modify configuration & \ref{sec:response} \\
\bottomrule
\end{tabular}
\end{table}

\end{document}